\documentclass[linenumbers]{aastex631}
\usepackage{placeins}

\newcommand{\gr}{{$\gamma$-ray}}

\shortauthors{IceCube collaboration}
\graphicspath{{./}{figures/}}

\begin{document}
\nolinenumbers
\title{Searches for Neutrinos from LHAASO ultra-high-energy \gr\ sources using the IceCube Neutrino Observatory}

\affiliation{III. Physikalisches Institut, RWTH Aachen University, D-52056 Aachen, Germany}
\affiliation{Department of Physics, University of Adelaide, Adelaide, 5005, Australia}
\affiliation{Dept. of Physics and Astronomy, University of Alaska Anchorage, 3211 Providence Dr., Anchorage, AK 99508, USA}
\affiliation{Dept. of Physics, University of Texas at Arlington, 502 Yates St., Science Hall Rm 108, Box 19059, Arlington, TX 76019, USA}
\affiliation{CTSPS, Clark-Atlanta University, Atlanta, GA 30314, USA}
\affiliation{School of Physics and Center for Relativistic Astrophysics, Georgia Institute of Technology, Atlanta, GA 30332, USA}
\affiliation{Dept. of Physics, Southern University, Baton Rouge, LA 70813, USA}
\affiliation{Dept. of Physics, University of California, Berkeley, CA 94720, USA}
\affiliation{Lawrence Berkeley National Laboratory, Berkeley, CA 94720, USA}
\affiliation{Institut f{\"u}r Physik, Humboldt-Universit{\"a}t zu Berlin, D-12489 Berlin, Germany}
\affiliation{Fakult{\"a}t f{\"u}r Physik {\&} Astronomie, Ruhr-Universit{\"a}t Bochum, D-44780 Bochum, Germany}
\affiliation{Universit{\'e} Libre de Bruxelles, Science Faculty CP230, B-1050 Brussels, Belgium}
\affiliation{Vrije Universiteit Brussel (VUB), Dienst ELEM, B-1050 Brussels, Belgium}
\affiliation{Department of Physics and Laboratory for Particle Physics and Cosmology, Harvard University, Cambridge, MA 02138, USA}
\affiliation{Dept. of Physics, Massachusetts Institute of Technology, Cambridge, MA 02139, USA}
\affiliation{Dept. of Physics and The International Center for Hadron Astrophysics, Chiba University, Chiba 263-8522, Japan}
\affiliation{Department of Physics, Loyola University Chicago, Chicago, IL 60660, USA}
\affiliation{Dept. of Physics and Astronomy, University of Canterbury, Private Bag 4800, Christchurch, New Zealand}
\affiliation{Dept. of Physics, University of Maryland, College Park, MD 20742, USA}
\affiliation{Dept. of Astronomy, Ohio State University, Columbus, OH 43210, USA}
\affiliation{Dept. of Physics and Center for Cosmology and Astro-Particle Physics, Ohio State University, Columbus, OH 43210, USA}
\affiliation{Niels Bohr Institute, University of Copenhagen, DK-2100 Copenhagen, Denmark}
\affiliation{Dept. of Physics, TU Dortmund University, D-44221 Dortmund, Germany}
\affiliation{Dept. of Physics and Astronomy, Michigan State University, East Lansing, MI 48824, USA}
\affiliation{Dept. of Physics, University of Alberta, Edmonton, Alberta, Canada T6G 2E1}
\affiliation{Erlangen Centre for Astroparticle Physics, Friedrich-Alexander-Universit{\"a}t Erlangen-N{\"u}rnberg, D-91058 Erlangen, Germany}
\affiliation{Physik-department, Technische Universit{\"a}t M{\"u}nchen, D-85748 Garching, Germany}
\affiliation{D{\'e}partement de physique nucl{\'e}aire et corpusculaire, Universit{\'e} de Gen{\`e}ve, CH-1211 Gen{\`e}ve, Switzerland}
\affiliation{Dept. of Physics and Astronomy, University of Gent, B-9000 Gent, Belgium}
\affiliation{Dept. of Physics and Astronomy, University of California, Irvine, CA 92697, USA}
\affiliation{Karlsruhe Institute of Technology, Institute for Astroparticle Physics, D-76021 Karlsruhe, Germany }
\affiliation{Karlsruhe Institute of Technology, Institute of Experimental Particle Physics, D-76021 Karlsruhe, Germany }
\affiliation{Dept. of Physics, Engineering Physics, and Astronomy, Queen's University, Kingston, ON K7L 3N6, Canada}
\affiliation{Department of Physics {\&} Astronomy, University of Nevada, Las Vegas, NV, 89154, USA}
\affiliation{Nevada Center for Astrophysics, University of Nevada, Las Vegas, NV 89154, USA}
\affiliation{Dept. of Physics and Astronomy, University of Kansas, Lawrence, KS 66045, USA}
\affiliation{Department of Physics and Astronomy, UCLA, Los Angeles, CA 90095, USA}
\affiliation{Centre for Cosmology, Particle Physics and Phenomenology - CP3, Universit{\'e} catholique de Louvain, Louvain-la-Neuve, Belgium}
\affiliation{Department of Physics, Mercer University, Macon, GA 31207-0001, USA}
\affiliation{Dept. of Astronomy, University of Wisconsin{\textendash}Madison, Madison, WI 53706, USA}
\affiliation{Dept. of Physics and Wisconsin IceCube Particle Astrophysics Center, University of Wisconsin{\textendash}Madison, Madison, WI 53706, USA}
\affiliation{Institute of Physics, University of Mainz, Staudinger Weg 7, D-55099 Mainz, Germany}
\affiliation{Department of Physics, Marquette University, Milwaukee, WI, 53201, USA}
\affiliation{Institut f{\"u}r Kernphysik, Westf{\"a}lische Wilhelms-Universit{\"a}t M{\"u}nster, D-48149 M{\"u}nster, Germany}
\affiliation{Bartol Research Institute and Dept. of Physics and Astronomy, University of Delaware, Newark, DE 19716, USA}
\affiliation{Dept. of Physics, Yale University, New Haven, CT 06520, USA}
\affiliation{Columbia Astrophysics and Nevis Laboratories, Columbia University, New York, NY 10027, USA}
\affiliation{Dept. of Physics, University of Oxford, Parks Road, Oxford OX1 3PU, UK}
\affiliation{Dipartimento di Fisica e Astronomia Galileo Galilei, Universit{\`a} Degli Studi di Padova, 35122 Padova PD, Italy}
\affiliation{Dept. of Physics, Drexel University, 3141 Chestnut Street, Philadelphia, PA 19104, USA}
\affiliation{Physics Department, South Dakota School of Mines and Technology, Rapid City, SD 57701, USA}
\affiliation{Dept. of Physics, University of Wisconsin, River Falls, WI 54022, USA}
\affiliation{Dept. of Physics and Astronomy, University of Rochester, Rochester, NY 14627, USA}
\affiliation{Department of Physics and Astronomy, University of Utah, Salt Lake City, UT 84112, USA}
\affiliation{Tsung-Dao Lee Institute, Shanghai Jiao Tong University, Shanghai 201210, China}
\affiliation{Oskar Klein Centre and Dept. of Physics, Stockholm University, SE-10691 Stockholm, Sweden}
\affiliation{Dept. of Physics and Astronomy, Stony Brook University, Stony Brook, NY 11794-3800, USA}
\affiliation{Dept. of Physics, Sungkyunkwan University, Suwon 16419, Korea}
\affiliation{Institute of Physics, Academia Sinica, Taipei, 11529, Taiwan}
\affiliation{Dept. of Physics and Astronomy, University of Alabama, Tuscaloosa, AL 35487, USA}
\affiliation{Dept. of Astronomy and Astrophysics, Pennsylvania State University, University Park, PA 16802, USA}
\affiliation{Dept. of Physics, Pennsylvania State University, University Park, PA 16802, USA}
\affiliation{Dept. of Physics and Astronomy, Uppsala University, Box 516, S-75120 Uppsala, Sweden}
\affiliation{Dept. of Physics, University of Wuppertal, D-42119 Wuppertal, Germany}
\affiliation{DESY, D-15738 Zeuthen, Germany}

\author[0000-0001-6141-4205]{R. Abbasi}
\affiliation{Department of Physics, Loyola University Chicago, Chicago, IL 60660, USA}

\author[0000-0001-8952-588X]{M. Ackermann}
\affiliation{DESY, D-15738 Zeuthen, Germany}

\author{J. Adams}
\affiliation{Dept. of Physics and Astronomy, University of Canterbury, Private Bag 4800, Christchurch, New Zealand}

\author{N. Aggarwal}
\affiliation{Dept. of Physics, University of Alberta, Edmonton, Alberta, Canada T6G 2E1}

\author[0000-0003-2252-9514]{J. A. Aguilar}
\affiliation{Universit{\'e} Libre de Bruxelles, Science Faculty CP230, B-1050 Brussels, Belgium}

\author[0000-0003-0709-5631]{M. Ahlers}
\affiliation{Niels Bohr Institute, University of Copenhagen, DK-2100 Copenhagen, Denmark}

\author[0000-0002-9534-9189]{J.M. Alameddine}
\affiliation{Dept. of Physics, TU Dortmund University, D-44221 Dortmund, Germany}

\author{A. A. Alves Jr.}
\affiliation{Karlsruhe Institute of Technology, Institute for Astroparticle Physics, D-76021 Karlsruhe, Germany }

\author{N. M. Amin}
\affiliation{Bartol Research Institute and Dept. of Physics and Astronomy, University of Delaware, Newark, DE 19716, USA}

\author{K. Andeen}
\affiliation{Department of Physics, Marquette University, Milwaukee, WI, 53201, USA}

% ...

\date{\today}

\collaboration{389}{IceCube Collaboration}



\begin{abstract}
Galactic PeVatrons are Galactic sources theorized to accelerate cosmic rays up to PeV in energy. The accelerated cosmic rays are expected to interact hadronically with nearby ambient gas or the interstellar medium, resulting in \gr s and neutrinos. Recently, the Large High Altitude Air Shower Observatory (LHAASO) identified 12 \gr\ sources with emissions above 100 TeV, making them candidates for PeV cosmic-ray accelerators (PeVatrons). While at these high energies the Klein-Nishina effect suppresses exponentially leptonic emission from Galactic sources, evidence for neutrino emission would unequivocally confirm hadronic acceleration. Here, we present the results of a search for neutrinos from these \gr\ sources and stacking searches testing for excess neutrino emission from all 12 sources as well as their subcatalogs of supernova remnants and pulsar wind nebulae with 11 years of track events from the IceCube Neutrino Observatory. No significant emissions were found. Based on the resulting limits, we place constraints on the fraction of \gr\ flux originating from the hadronic processes in the Crab Nebula and LHAASO\,J2226+6057.

\end{abstract}



\section{Introduction} \label{sec:intro}

The origin of the high-energy cosmic rays (CRs) is a longstanding question in particle astrophysics. 
The cosmic-ray spectrum is typically parametrized with an E$^{-2.7}$ power law up to its ``knee'', ${\rm E}_{\rm CR}\sim 3$~PeV, above which the spectrum softens \citep{Abbasi2018}. 
The CRs with energy up to the ``knee'' are generally believed to be confined and accelerated in the Milky Way, implying the presence of PeV accelerators known as PeVatrons that accelerate the particles to energy $>1$~PeV within our Galaxy. 
In addition to propelling CRs to PeV energies, Galactic PeVatrons are expected to emit \gr s and neutrinos through associate pion decays \citep{Ahlers2016}. 
High energy \gr s may also originate from leptonic processes like inverse Compton scattering and bremsstrahlung; however, at ${\rm E}_{\gamma} \gtrsim 50$~TeV, inverse Compton scattering is inefficient due to Klein-Nishina suppression \citep{Cristofari2021,Hinton2009,Moderski2005}. 
As inverse Compton scattering is the primary mechanism for electrons to emit \gr s \citep{Blumenthal1970}, \gr\ emission from leptonic processes is dramatically suppressed. 
Ultra-high-energy (UHE, $>100$~TeV) \gr\ sources, especially those with observed spectra following a simple power law to at least tens of TeV, are suspected to be indicators of PeVatrons. 
Testify that the acceleration of hadrons is taking place for those TeV sources is of importance to pinpoint the PeV CR sources. 


Supernova remnants (SNRs) are widely considered candidate Galactic PeVatrons. 
SNRs are generally believed to be the primary candidates of hadronic sources in the Galaxy, with particles being accelerated by diffusive shock acceleration (DSA) in expanding shocks. According to the SNR hypothesis, 10\% of the total explosion energy of the supernova progenitor is converted into CRs, compatible with the observed intensity of CRs and their estimated confinement time in the Galaxy \citep{1978ApJ...221L..29B}. The DSA mechanism and strong magnetic field amplification close to the shocks in SNRs also can explain the observed power-law spectrum of CRs \citep{Cristofari2018}. Apart from that, molecular clouds located within~$100$~parsec of the SNRs could be effective sites for $>$~TeV \gr\ production \citep{Gabici2007}. The highest-energy particles would quickly escape SNRs and diffuse faster in the interstellar medium (ISM), arriving at nearby molecular clouds earlier than the lower energy CRs and producing neutrinos and \gr s there. 
CR might also be accelerated in young massive star clusters (YMCs) and pulsar wind nebulae (PWNe). 
Although the \gr\ emission from PWNe is widely expected to be leptonic, several studies suggest that the presence of relativistic hadrons in the pulsar winds might also be possible \citep{Dipalma2017}. 
From the Hillas criterion, the magnetic fields of PWNe could effectively trap particles with energy ${\rm E} \lesssim 3$~PeV  \citep{Hillas1984}. 
Therefore, a large amount of energy of PWNe could be stored in relativistic protons if they are continuously produced. 
In hadronic scenarios, CRs might be accelerated at the termination shocks \citep{Dipalma2017} or by the shocks of SNRs which confined the diffuse nebulae \citep{Ohira2018}. 


\citet{Albert2021pev} reported that the TeV \gr\ spectrum of HAWC\,J1825-134 has no indication for a cut-off up to 200 TeV, implying that the source might be a possible PeVatron. 
HAWC\,J1825-134 may be associated with a giant molecular cloud, and the YMC located inside the molecular cloud is thought to accelerate CRs to PeV energies. 
Evidence of a PeVatron at the center of the Galaxy has also been reported by the HESS Collaboration by observing  \gr s from collisions of a young population of CRs with close by molecular clouds, with a hard \gr\ spectrum observed up to ${\rm E}_{\gamma} \sim 40$ TeV \citep{HESS2016}.  
Detection of astrophysical neutrinos (${\rm E}_{\nu} \gtrsim 100$~TeV) would be a smoking gun for the existence of Galactic PeVatrons. 
The IceCube Neutrino Observatory has previously searched for Galactic neutrino emission from individual sources, stacked similar objects (PWNe, SNRs, etc.), and the Galactic plane \citep{Aartsen2017galactic,Aartsen2019Cascade,aartsen2020time,Aartsen2020PWN}. 
Joint analyses with data from the ANTARES Neutrino Telescope were also performed to search for neutrinos from the Galactic plane and the southern sky \citep{Albert2018galactic,Albert2020Southern}, exploiting the advantageous field of view of ANTARES and large volume and high statistics of IceCube. 
No significant Galactic neutrino sources were found from those searches, and the contribution of Galactic sources to the all-sky isotropic astrophysical neutrino flux has been constrained to $\leq 14\%$ \citep{Aartsen2017galactic}. 

Recently, the LHAASO Collaboration reported the detection of a dozen Galactic ${\rm E}_{\gamma} \geq 100$~TeV \gr\ sources with $> 7 \sigma$ detection \citep{cao2021ultrahigh}. 
This is the first LHAASO catalog and the largest list of UHE \gr\ sources published to date. 
Searching for neutrino emission from those LHAASO UHE sources will enable differentiation between leptonic and hadronic mechanisms that contribute to UHE \gr\ flux. 
In this letter, we perform a stacking analysis and a catalog search to probe the hadronic nature of these potential PeVatrons. 
We divide the LHAASO sources into 3 groups according to their possible association (all UHE sources, all possible PWNe, and all possible SNRs) in the stacking analysis, providing flux limits to constrain the total fraction of Galactic CRs that come from the UHE \gr\ SNRs and PWNe above 100~TeV. 
\hyperref[sec:icecube]{Section~\ref*{sec:icecube}} briefly introduces the IceCube Observatory, the LHAASO detector arrays, the neutrino dataset, and the source list of 12 UHE sources. 
Analysis methods are shown in \hyperref[sec:analysis]{Section~\ref*{sec:analysis}}. 
In \hyperref[sec:discuss]{Section~\ref*{sec:discuss}} and \hyperref[sec:conclude]{Section~\ref*{sec:conclude}}, we discuss and summarize the results. 

\section{IceCube and LHAASO detectors and datasets} \label{sec:icecube}
\subsection{IceCube neutrino sample} 
The IceCube Neutrino Observatory is a cubic kilometer
array of 5160 digital optical modules (DOMs) located in the geographic South Pole \citep{abbasi2009icecube}. These DOMs are arranged along 86 readout and support cables ("strings") that instrument the glacial ice at a depth between 1.45 to 2.45 km. Each DOM is equipped with a 10-inch photomultiplier tube 
in a pressure-resistant sphere along with the necessary electronics for read-out of signal \citep{abbasi2009icecube}. These DOMs are designed to observe the Cherenkov light induced by energetic charged particles traveling in ice. These energetic charged particles are produced by charged-current (CC) interaction or neutral-current (NC) interaction from neutrinos of all flavors \citep{ABBASI2010139}. The track-like (from CC interaction from muon neutrino) and cascade-like (from other NC and CC interaction) events are recorded by DOMs and analyzed in terms of the direction and energy of the primary neutrino. IceCube also includes DeepCore, a GeV-scale neutrino sub-detector \citep{2012APh....35..615A} and IceTop, an air shower array \citep{2013NIMPA.700..188A}, which are not used in this study.

The dataset used in this analysis comprises 11 years of IceCube track-like data from April 6, 2008 to May 29, 2020. Track-like events have a better angular resolution (a typical $\sim$ TeV neutrino has angular uncertainty $\sim 1^{\circ}$) compared to cascade-like events because of the long lever arm, providing better sensitivity to astrophysical point sources. The data taken from April 6, 2008 to May 13, 2011 are from the partially built detector with 40, 59 and 79 strings (IC40, IC59, and IC79) \citep{aartsen2020time} and the rest of the data are taken from the the full detector configuration of 86 strings \citep{aartsen2014searches}. The dominant source of background is atmospheric neutrinos and atmospheric muons originating from air showers induced by cosmic-ray interactions in the atmosphere. \hyperref[dataset]{Appendix~\ref*{dataset}} shows the details of the dataset used and the relevant energy range.

The dataset also incorporates the improved in-situ calibration of the single-photoelectron charge distributions \citep{aartsen2020situ} and other improvements in reconstruction, further boosting the sensitivity relative to the previous analyses \citep{aartsen2020time}. Sensitivity and 5$\sigma$ discovery potential of the dataset are shown in \hyperref[tab:fig1]{Figure~\ref*{tab:fig1}}.

\newpage

\subsection{LHAASO ultra-high-energy \gr\ sources} 

The Large High Altitude Air Shower Observatory (LHAASO) is a complex of extensive air shower detector arrays aims to study \gr s and cosmic rays. Located in the Sichuan province of China with an elevation of $4410~m$ a.s.l., LHAASO consists of the three components Kilometer Square Array (KM2A), Water Cherenkov Detector Array (WCDA), and Wield Field-of-view Cherenkov Telescope Array (WFCTA). LHAASO-KM2A focuses on Galactic \gr\ sources above $30$~TeV in the Northern sky and has been operating since December 2019. The KM2A comprises uniformly distributed electromagnetic particle detectors (ED) and muon detectors (MD) over $1.3~{\rm km}^{2}$. The vast area of scintillation counters together with the MD array would effectively suppress the background from CR-induced hadronic showers for \gr s above $\sim 50$~TeV, yielding a sensitivity of the full KM2A approaching $\sim 1\%$ for a Crab-like source with a 1-year observation at 100 TeV \citep{Liu2016,Cui2014}. The angular and energy resolution of the KM2A are $0.4^\circ$ and $28\%$ at 30~TeV, respectively. We note that at 1 PeV, the angular resolution of the KM2A would reach $0.2^\circ$ \citep{Cao2022}. 

The half-completed LHAASO-KM2A has already detected twelve UHE \gr\ sources above 100 TeV with statistical significance $\geq 7\sigma$, from December 2019 to October 2021 \citep[][Table~\ref{tab:Table1}]{cao2021ultrahigh}. All these potential PeVatrons appear to be located in our Galaxy. 
Among these twelve sources, three sources (eHWCJ1825-134, eHWCJ1907+063, and eHWCJ2019+368) are also in the HAWC's high energy source list \citep{eHAWCcatalog} with conclusive evidences of $\geq 100$~TeV emission. 
Except for LHAASO J0534+2202, which is the Crab Nebula, all the other LHAASO UHE sources have no firm association, with several known TeV counterparts nearby. 
Table 2 in \citet{cao2021ultrahigh} lists all the possible TeV associations close to these twelve UHE sources, suggesting that majority of them might be associated with PWNe or SNRs. 

\section{Analysis Method} \label{sec:analysis}

\subsection{Catalog search}

A maximum likelihood ratio test is performed following the technique described in \cite{abbasi2011time} and \cite {braun2008methods}. The same method was also used in the previous IceCube 10 years time-integrated point source search \citep{aartsen2020time}. The first LHAASO catalog consists of 12 UHE $\gamma$-ray sources detected above 100 TeV with $>7\sigma$  \citep{cao2021ultrahigh}. Only two of the LHAASO sources (LHAASO J0534+2202 and LHAASO J2108+5157) are considered point sources with regard to LHAASO's angular resolution. For the remaining ten sources, significant angular extensions in $\gamma$-ray are detected by LHAASO and a 0.3 degree Gaussian emission template is used to measure the $\gamma$-ray flux. Hence, a 0.3 degree Gaussian extent is assumed for the remaining ten sources to match assumptions used in the LHAASO analysis \citep{cao2021ultrahigh}. We model the spatial signal likelihood as a 2D Gaussian in tangent plane coordinates which is valid for a sufficiently small angular extension. The angular uncertainty of each event is obtained directly from the reconstruction. Source extensions and estimated angular uncertainties for events are added in quadrature when estimating spatial likelihoods for each event. We assume a power-law spectrum with spectral index $\Gamma$ for the signal hypothesis. The normalization and spectral index $1<\Gamma<4$ of the signal flux are optimized in a maximum likelihood analysis. A detailed description of the likelihood can be found in \hyperref[likelihood]{Appendix~\ref*{likelihood}}. 

\hyperref[tab:fig1]{Figure \ref*{tab:fig1}} shows the sensitivity and 5$\sigma$ discovery potential as a function of declination of this search method for $\Gamma = 2$ and $\Gamma = 3$. Sensitivity is the neutrino flux required for 90\% of the trials to yield a p-value less than 0.5. The 5~$\sigma$ discovery potential is the neutrino flux required for 50\% of trials with simulated signal to yield a p-value less than $2.87\times10^{-7}$(5~$\sigma$ significance in one-tail Gaussian test). In our analyses, we use 50~TeV as the pivot energy when calculating the neutrino flux, given that 100~TeV \gr s from pion decay would be accompanied by neutrinos at 50~TeV (\citet{2014PhRvD..90b3010A}, see \hyperref[convert]{Appendix~\ref*{convert}} for more detail). 

\subsection{Stacking search}
A total of six stacking analyses are performed combing three catalogues - all LHAASO sources, sources potentially associated with SNRs, and sources potentially associated with PWNe - and two weighting schemes - equal neutrino flux at Earth and neutrino flux proportional to the $\gamma$-ray flux at 100 TeV reported by LHAASO \citep{cao2021ultrahigh}. 
The source hypothesis in the stacking analysis is the same as in the catalog search, with sources considered to have 0.3 degree Gaussian extension except for LHAASO J0534+2202 and LHAASO J2108+5157. 
According to the association proposed by \citet{cao2021ultrahigh}, nine sources are included in the PWN stacking search, and six sources are included in the SNR stacking search. There are four gamma-ray sources in common between the SNR and PWN stacking searches.  \hyperref[tab:Table1]{Table \ref*{tab:Table1}} shows the source type of the potential association of the LHAASO sources. We assume that all the sources in each stacking scheme have a universal unbroken power-law energy spectrum with the same spectral index. The spectral index and the total number of signal neutrinos are fitted as free parameters during the maximization of the likelihood. For the equal-weight cases, the relative weight of each source is $\frac{1}{N}$ where N is the number of sources in the catalog. For the flux-weighted cases, the relative weight of each source i is $\frac{\phi_i}{\sum_j \phi_j}$ where $\phi_i$ is the LHAASO $\gamma$-ray flux at 100 TeV \citep{cao2021ultrahigh}. We generate $10^{5}$ background trials by scrambling the right ascension coordinates of the track sample. The stacking search is performed on each background trial, identifying the source with the lowest pre-trial p-value in each. The distribution of these p-values is used to correct for the look-elsewhere effect. The post-trial p-value is computed by comparing the pre-trial p-value obtained from the real data and the distribution of the lowest pre-trial p-values \citep{aartsen2020time}. 

\section{Results and discussion} \label{sec:discuss}
\subsection{Catalog search and Stacking search results}
\begin{figure}[h!]
  \centering
  \begin{minipage}[b]{0.49\textwidth}
    \centering
    \includegraphics[width=\textwidth]{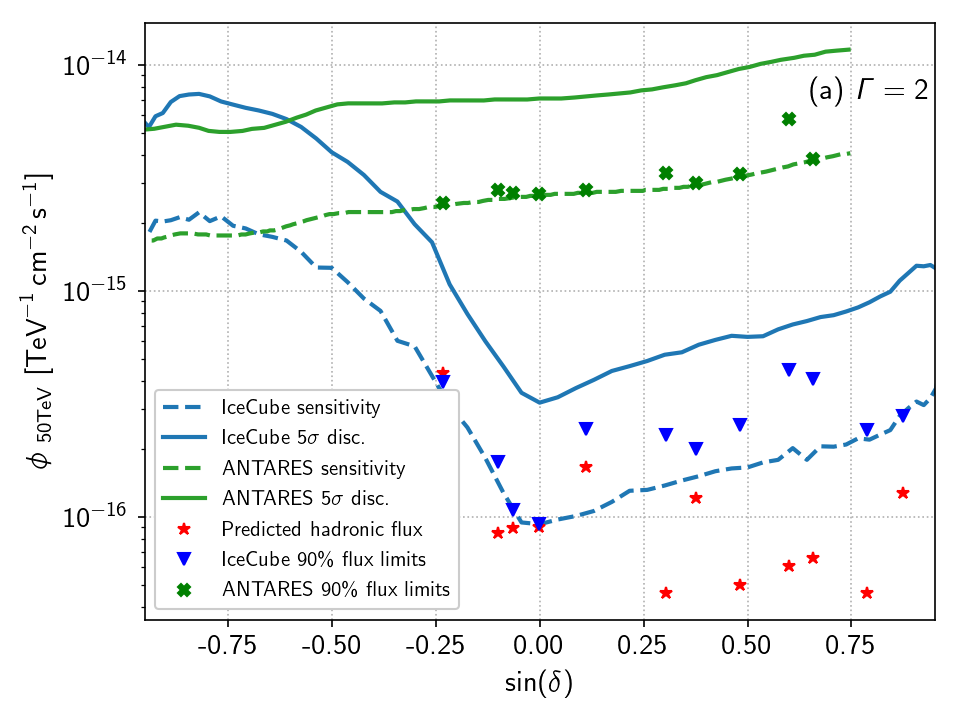}
  \end{minipage}
  \hfill
  \begin{minipage}[b]{0.49\textwidth}
    \centering
    \includegraphics[width=\textwidth]{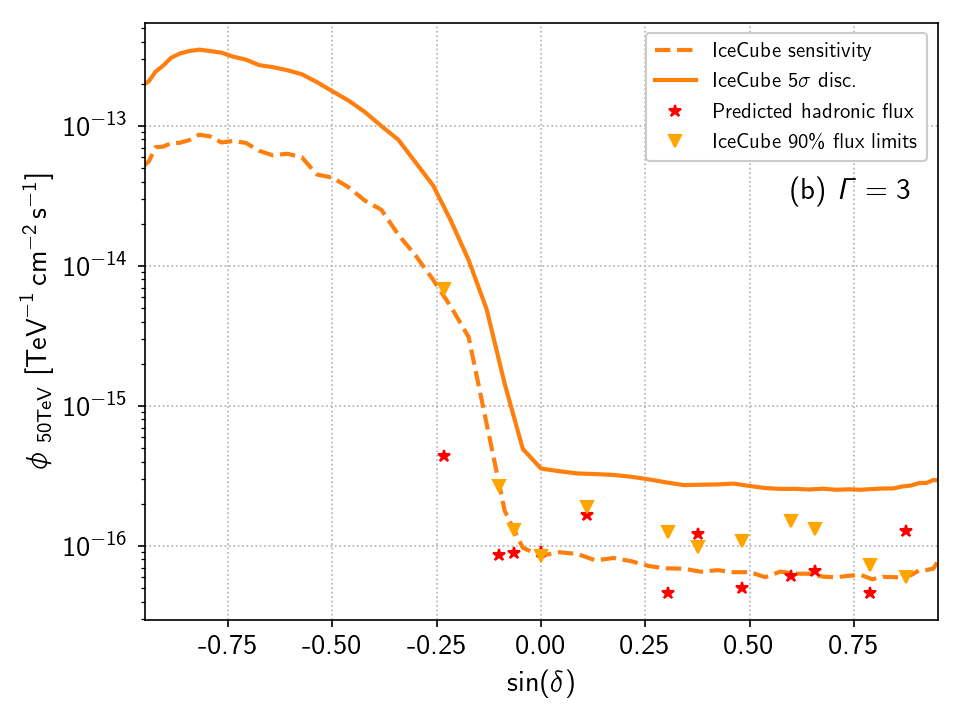}
  \end{minipage}
  \caption{The blue solid and blue dashed lines are the per-source IceCube 5$\sigma$ discovery potential and sensitivity for 0.3 degree extension and $\Gamma = 2$ spectrum (a), and the orange solid and dashed lines are those for 0.3 degree extension and $\Gamma = 3$ spectrum (b) of this analysis. The green solid and green dashed lines on the left plot are the 5$\sigma$ discovery potential and sensitivity of a $E^{-2}$ spectrum of ANTARES point source analysis \citep{illuminati2021searches}. The red stars represent the neutrino flux predicted from the LHAASO measurement if photon flux is assumed to be 100\% of hadronic origin \citep{2014PhRvD..90b3010A}. The blue and orange triangles represent the 90\% flux limits obtained from this analysis for $E^{-2}$ and $E^{-3}$ spectrum, respectively. The green crosses are the ANTARES 90\% flux limits \citep{illuminati2021searches}.
}
\label{tab:fig1}
\end{figure}

\hyperref[tab:Table1]{Table \ref*{tab:Table1}} shows the result of the catalog search, including the best-fit number of neutrino events, best-fit power law index and pre-trial p-value. The 90\% confidence interval (C.I.) flux upper limit for each source at 50 TeV assuming a spectral index $\Gamma = 2$ is shown in the last column. We find no significant correlation between the observed neutrinos and the newly identified LHAASO UHE (${\rm E}_{\gamma} > 100$~TeV) sources in our searches. \hyperref[tab:fig1]{Figure \ref*{tab:fig1}} shows 90\% flux limits for $\Gamma = 2$ and $\Gamma = 3$, comparing with the sensitivity from IceCube's 11 years dataset and ANTARES's 13 years dataset \citep{illuminati2021searches}. \hyperref[convert]{Appendix~\ref*{convert}} explains the detail of the estimation of the neutrino flux from the observed gamma-ray flux \citep{2014PhRvD..90b3010A}. Here we assume that the full \gr\ flux is produced via proton-proton interaction. 

\begin{table}[h!]
\centering
\begin{tabular}{|cccccccccc|}

\hline
\multicolumn{1}{|c|}{Source} & \multicolumn{1}{c|}{R.A.} & \multicolumn{1}{c|}{Dec} & \multicolumn{1}{c|}{$\gamma$-ray flux[CU]} & \multicolumn{1}{c|}{Possible association} & \multicolumn{1}{c|}{$n_s$} & \multicolumn{1}{c|}{$\Gamma$} & \multicolumn{1}{c|}{TS} & \multicolumn{1}{c|}{pre-trial p-value} & \multicolumn{1}{c|}{$\phi_{90\%}$ } \\ \hline

LHAASOJ1825-1326                                                            & 276.45                  & $-$13.45          &3.57         & \textbf{PWN}                       & 1.00                    & 3.33                       & 0.02                    & 0.42                 & 4.0                            \\
LHAASOJ1839-0545                                                            & 279.95                  & $-$5.75         &0.7           & \textbf{PWN}                       & 9.34                    & 3.12                       & 1.43                    & 0.46                 &1.8                            \\
LHAASOJ1843-0338                                                            & 280.75                  & $-$3.65          &0.73          & \textbf{SNR}                       & 0.00                    & ---                       & 0.00                    & 1.0                 & 0.99                            \\
LHAASOJ1849-0003                                                            & 282.35                  & $-$0.05          &0.74          & \textbf{PWN}/YMC                       & 0.00                    & ---                       & 0.00                    & 1.0                 & 0.90                            \\
LHAASOJ1908+0621                                                            & 287.05                  & 6.35             &1.36        & \textbf{SNR/PWN}                   & 6.83                    & 2.11                       & 4.06                    & 0.046                 & 2.5                            \\
LHAASOJ1929+1745                                                            & 292.25                  & 17.75           &0.38         & \textbf{SNR/PWN}                   & 16.0                   & 2.63                       & 1.34                    & 0.18                 &2.3                            \\
LHAASOJ0534+2202                                                            & 83.55                   & 22.05          &1.0          & \textbf{PWN}                       & 14.0                   & 4.0                       & 1.22                    & 0.19                  &2.0                           \\
LHAASOJ1956+2845                                                            & 299.05                  & 28.75         &0.41           & \textbf{SNR}                       & 17.6                   & 3.05                       & 1.16                    & 0.21                &2.5                             \\
LHAASOJ2018+3651                                                            & 304.75                  & 36.85         &0.5           & \textbf{PWN}/YMC                   & 18.7                   & 2.67                       & 3.62                    & 0.045                  &4.5                           \\
LHAASOJ2032+4102                                                            & 308.05                  & 41.05         &0.54           & \textbf{SNR/PWN}/YMC                   & 24.8                   & 3.98                       & 2.81                    & 0.075                 &4.1                            \\
LHAASOJ2108+5157                                                            & 317.15                  & 51.95         &0.38           & -                         & 10.6                   & 2.96                       & 0.84                    & 0.26                  &2.4                           \\
LHAASOJ2226+6057                                                            & 336.75                  & 60.95         &1.05           & \textbf{SNR/PWN}                   & 0.00                    & ---                        & 0.00                    & 1.0                 &2.9                           \\ \hline
\end{tabular}

\caption{Table of best-fit parameters with corresponding test statistic (TS) and p-value of catalog search. The $\gamma$-ray flux is the $\gamma$-ray flux of the source at 100 TeV in Crab units (CU) ($6.1 \times 10^{-17}~{\rm TeV}^{-1}~{\rm cm}^{-2}~{\rm s}^{-1}$) reported by LHAASO \citep{cao2021ultrahigh}. The neutrino 90\% C.I. flux upper-limit ($\phi_{90\%}$) is parameterized as: $\frac{dN_{\nu_{\mu} +\nu_{\bar{\mu}}}}{dE_\nu} = \phi_{90\%} \cdot (\frac{E_\nu}{50 {\rm TeV}})^{-2} \times 10^{-16}~{\rm TeV}^{-1}~{\rm cm}^{-2}~{\rm s}^{-1}$. The relevant energy range of the flux upper limit can be found in \hyperref[dataset]{Appendix C}. The two smallest pre-trial p-values of 0.046 and 0.045 observed for two sources correspond to a post-trial p-value of 0.42 with the assumption that those 12 sources are independent given their large separation.}
\label{tab:Table1}
\end{table}

The two lowest p-value sources are LHAASO J1908+0621 and LHAASO J2018+3651. 
LHAASO J1908+0621 is thought to be the UHE counterpart of MGRO J1908+06, which is also the lowest p-value Galactic source in the source list of the IceCube 10-year point-source search \citep{aartsen2020time}. 
LHAASO J2018+3651 might be associated with MGRO J2019+37, and an HII region, Sh $2-104$, is located to the west. 
The region Sh $2-104$ comprises several YMCs with diffuse X-ray emission probably from the colliding winds of young stars \citep{Gotthelf2016}. 

Ten of the LHAASO UHE sources with $\delta < 42^\circ$ are in the source list for the ANTARES 13-year point-source search \citep{illuminati2021searches}. 
LHAASO J2018+3651 is the only LHAASO source with reported pre-trial p-value of 0.14. 
LHAASO J1908+0621, LHAASO J1929+1745, and LHAASO J2032+4102 have p-values $\lesssim 0.06$ in the searches of \citet{Huang2021} with 10 years of IceCube track events \citep{IceCube10yrs_dataset}. 
The pre-trial p-values of LHAASO J1929+1745 and LHAASO J2032+4102 are $\sim 0.03$ and $\sim 0.04$ with an extension of 0.6 degree, while that of LHAASO J1908+0621 is $\sim 0.06$ for the point-source hypothesis. 
LHAASO J2032+4132 is the third significant source in our catalog search with a pre-trial p-value of 0.075. 
The differences between the significance level for LHAASO J2018+3651 and LHAASO J1929+1745 in our analysis and that in \citet{Huang2021} might result from the differences in datasets' livetime and in track reconstruction approaches or from the limitations of the public dataset.

\begin{figure}[!h]
  \centering
  \begin{minipage}[b]{0.49\textwidth}
  \centering
    \includegraphics[width=\textwidth]{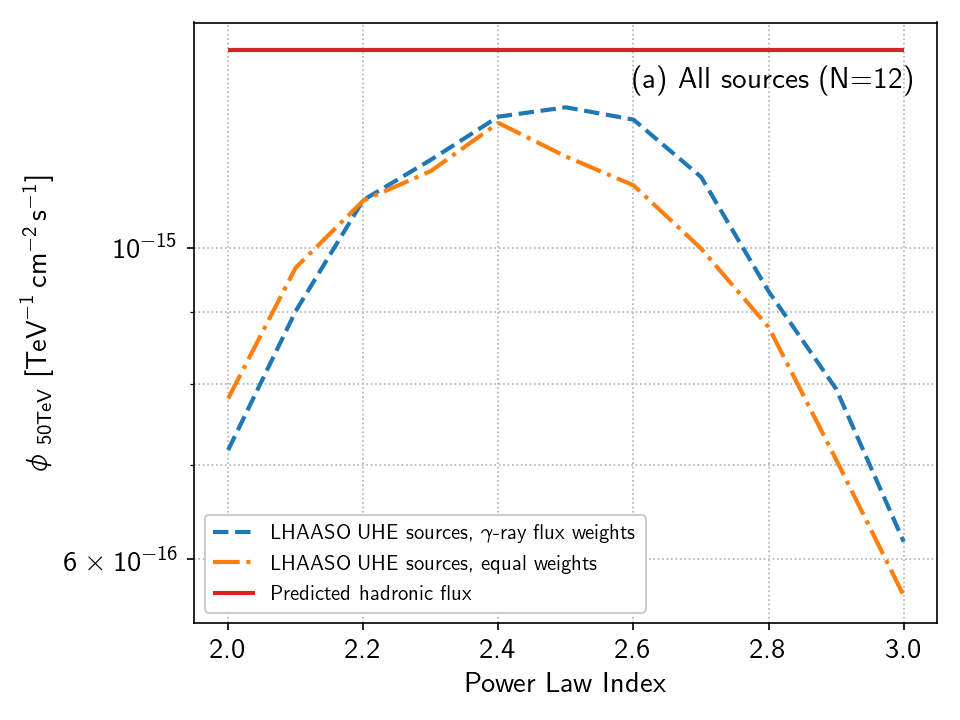}
  \end{minipage}
  \hfill
  \begin{minipage}[b]{0.49\textwidth}
  \centering
    \includegraphics[width=\textwidth]{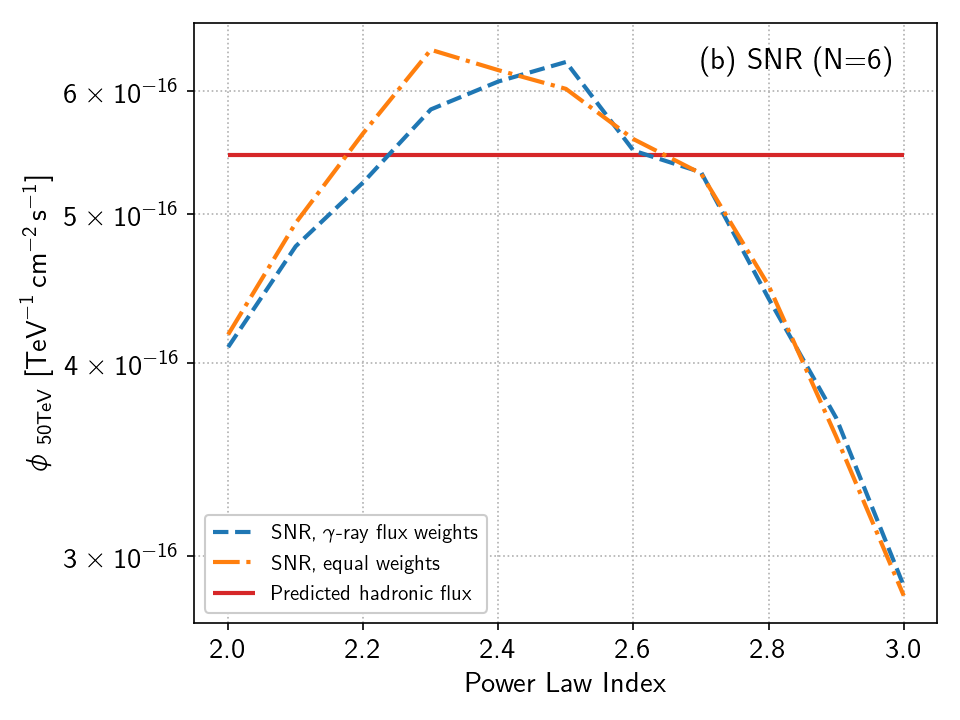}
  \end{minipage}
  \begin{minipage}[b]{0.49\textwidth}
  \centering
    \includegraphics[width=\textwidth]{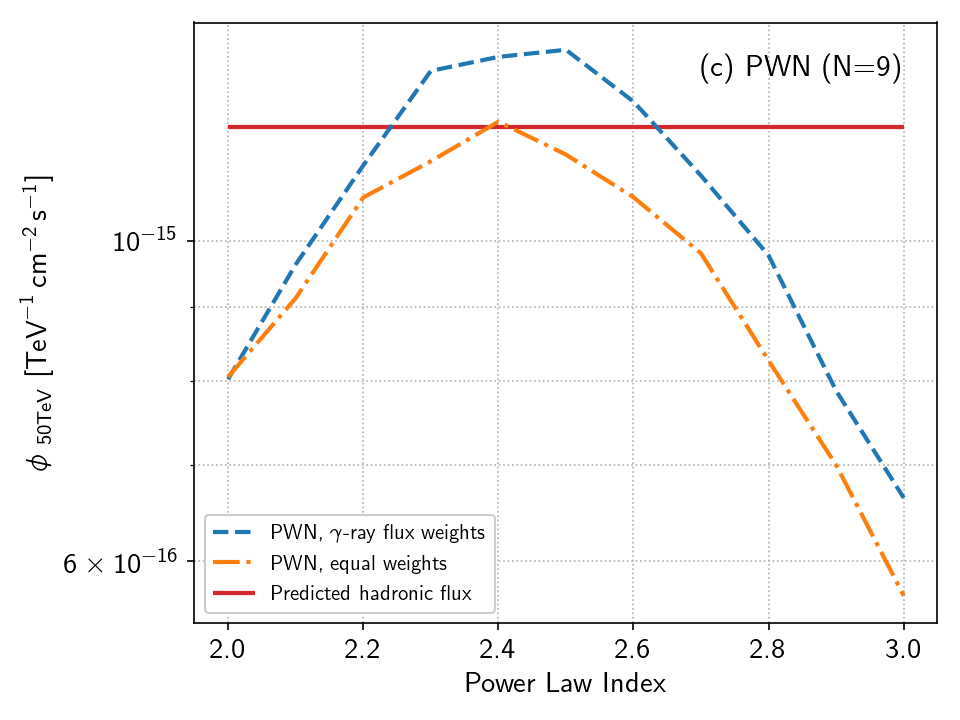}
  \end{minipage}
  \caption{Our 90\% flux upper limits on neutrino emission at 50 TeV from (a) all 12 LHAASO sources, (b) sources that could be associated with SNRs, and (c) sources that could be associated with PWNe. The orange line represents the equal-weight case and the blue line represents the flux-weighted case. The hadronic flux prediction assumes that the flux measured by LHAASO is of purely hadronic origin.} \label{tab:fig2}
\end{figure}

\hyperref[tab:table2]{Table \ref*{tab:table2}} shows the result of the stacking searches, including the best-fit total number of signal events, best-fit power law index, and pre-trial p-value for each stacking scheme. The 90\% C.I. flux upper limit at 50 TeV for each tested weighting scheme assuming a spectral index of 2 is shown in the last column. 
No significant emission is found from all 6 stacking schemes, with the lowest p-value results found from all possible LHAASO PWNe and all LHAASO UHE sources weighted equally. The 90\% flux upper limits for each weighting schemes tested as a function of spectral index is shown in \hyperref[tab:fig2]{Figure \ref*{tab:fig2}}. 

\begin{table}[h!]
\centering
\begin{tabular}{|ccccccc|}
\hline
\multicolumn{1}{|c|}{Sample } &\multicolumn{1}{|c|}{Weighting scheme} & \multicolumn{1}{c|}{$n_s$} & \multicolumn{1}{c|}{$\Gamma$} & \multicolumn{1}{c|}{TS} & \multicolumn{1}{c|}{Pre-trial p-value} & \multicolumn{1}{c|}{$\phi_{90\%}$} \\ \hline
All sources & flux-weighted                                                                     & 38.7               & 2.58                     & 3.11                  & 0.068                               & 7.2           \\
All sources & equal-weighted                                                                    & 79.7                 & 2.81                     & 4.83                  & 0.022                                 & 7.8           \\
SNR & flux-weighted                                                                     & 8.74                  & 2.21                     & 1.85                  & 0.14                                 & 4.1           \\
SNR & equal-weighted                                                                    & 22.3                 & 2.56                     & 1.42                  & 0.16                                 & 4.2           \\
PWN & flux-weighted                                                                     & 33.4                 & 2.46                     & 4.15                  & 0.042                                 & 8.0           \\
PWN & equal-weighted                                                                    & 59.2                 & 2.62                     & 5.42                  & 0.022                                 & 8.1           \\ \hline
\end{tabular}
\caption{Table of best-fit parameters for the stacking analyses. The 90\% flux upper-limit ($\phi_{90\%}$) is parameterized as
:$\frac{dN_{\nu_{\mu} +\nu_{\bar{\mu}}}}{dE_\nu} = \phi_{90\%} \cdot (\frac{E_\nu}{50 {\rm TeV}})^{-2} \times 10^{-16}~{\rm TeV}^{-1}~{\rm cm}^{-2}~{\rm s}^{-1}$. The smallest pre-trial p-value of 0.022 observed in the two stacking hypotheses corresponds to a post-trial p-value of 0.06.} \label{tab:table2}
\end{table}

\subsection{Hadronic constrains}
In this section, we try to constrain the fraction of \gr\ emission from hadronic processes in the LHAASO sources. 
To obtain meaningful hadronic constraints based on the non-observation of neutrinos, we look up the observed \gr\ indexes for LHAASO sources from the literature in order to estimate a more realistic \gr\ spectrum for the 90\% C.I. upper limits. 
Considering the observed TeV \gr\ spectra and comparing the theoretically predicted neutrino flux (the estimation detail is described in \hyperref[convert]{Appendix~\ref*{convert}}) with $90\%$ upper-limit flux calculated using the TeV spectral parameters, we are able to constrain the hadronic contributions to the \gr\ flux for three of the LHAASO UHE sources: LHAASO J0534+2202, LHAASO J1849-0003, and LHAASO J2226+6057. 
The hadronic constraints, neutrino upper limits, and expected neutrino fluxes for the three LHAASO sources are shown in Table~\ref{tab:table3}. 
We note that the constraints illustrated here are the best possible bounds we could obtain as we assume that the \gr\ emissions are completely hadronic. 
The limits would be weaker if part of the UHE \gr s from the sources originated from leptonic processes. 

\begin{table}[h!]
\centering
\begin{tabular}{|cccccccccc|}
\hline
\multicolumn{1}{|c|}{LHAASO name} & \multicolumn{1}{c|}{TeV counterpart} & \multicolumn{1}{c|}{Index / $\alpha$ } & \multicolumn{1}{c|}{$\beta$} & \multicolumn{1}{c|}{Energy range} & \multicolumn{1}{c|}{Extension} & \multicolumn{1}{c|}{$\phi_{90\%}$} & \multicolumn{1}{c|}{$\phi_{\rm Predicted}$} & \multicolumn{1}{c|}{Constraints} & \multicolumn{1}{c|}{Refs}  \\ 
\multicolumn{1}{|c|}{} & \multicolumn{1}{c|}{} &\multicolumn{1}{c|}{} & \multicolumn{1}{c|}{} & \multicolumn{1}{c|}{[TeV]} & \multicolumn{1}{c|}{[deg.]} & \multicolumn{1}{c|}{} & \multicolumn{1}{c|}{} & \multicolumn{1}{c|}{} & \multicolumn{1}{c|}{} \\ \hline
J1849$-$0003  & HESSJ1849$-$000  & 1.99   & ---  & $0.4-100$ & 0.09  & 0.85 & 0.90 & $<$94\% & [a] \\
J0534+2202    & Crab Nebula.    & 3.12   & ---  & $10-1600$ & 0.00  & 0.70 & 1.2 & $<$59\% & [b] \\ 
              &                  & 2.79   & 0.1  & $1-177$ & 0.00  & 1.0 & 1.2 & $<$84\% & [c] \\
J2226+6057    & SNR G106.3+02.7  & 3.01   & ---  & $20-500$ & 0.36 & 0.60 & 1.3 & $<$47\% & [d] \\ 
              &                  & 1.56   & 0.88 & $20-500$ & 0.36 & 2.1 & 1.3  & --- & [d] \\ 
\hline
\end{tabular}
\caption{Table of TeV spectral parameters and the corresponding hadronic constraints, neutrino upper limits, and expected neutrino fluxes at 50 TeV. The parameter $\phi_{90\%}$ represents the neutrino $90\%$ C.I. flux limits parameterized as :$\frac{dN_{\nu_{\mu} +\nu_{\bar{\mu}}}}{dE_\nu} = \phi_{90\%} \cdot (\frac{E_\nu}{50 {\rm TeV}})^{{-\alpha}-\beta\cdot \log\frac{E_\nu}{50 {\rm TeV}}} \times 10^{-16}~{\rm TeV}^{-1}~{\rm cm}^{-2}~{\rm s}^{-1}$. 
The parameter $\phi_{\rm Predicted}$ is the predicted neutrino flux with the assumption of \gr\ s are entirely hadronic. Hadronic constraints correspond to the ratio $\phi_{90\%}/\phi_{\rm Predicted}$ at 50 TeV calculated with fixed spectral parameters. The TeV spectral and morphology information (columns $3-6$) is taken from [a] \citet{Huang2021}, [b] \citet{2021Sci...373..425L}, [c] \citet{Abeysekara2019hawccrab}, and [d] \citet{cao2021ultrahigh}. }
\label{tab:table3}
\end{table}

The strongest hadronic constraint is put on LHAASO J2226+6057 with a limit of $\lesssim 47\%$, with the assumption that the neutrino spectrum follows a simple power law with an index of $3.01$ reported by LHAASO. 
The UHE source may be associated with SNR G106.3+02.7, suggested as a potential PeVatron by \citet{Albert2020Boomerang} and \citet{ASgamma2021}. 
The observations of HAWC and Tibet AS$\gamma$ for the SNR indicate an unbroken hard TeV spectrum extends up to at least 100 TeV. 
\citet{ASgamma2021} suggested that the VHE protons might be accelerated and trapped in the SNR given the hard spectral index. 
On the other hand, the UHE \gr\ flux of J2226+6057 detected by LHAASO from 20 to 500 TeV is better described with log-parabola rather than a single power-law \citep{cao2021ultrahigh}. 
However, the hadronic component of J2226+6057's \gr\ flux cannot be constrained when applying the log-parabola model to calculate the neutrino flux limits. 

For LHAASO J0534+2202, which is the UHE counterpart of the Crab Nebula, we constrain the hadronic UHE \gr\ emission to $\lesssim 59\%$ of the total, assuming a power law of $\Gamma=3.12$ as reported by LHAASO. 
Applying the log-parabola model by HAWC, the hadronic constraint on the Crab is $\lesssim 84\%$, compared to $\lesssim 70\%$ reported previously by \citet{Huang2021}. 
Although a 1.12 PeV $\gamma$-induced shower was detected by LHAASO-KM2A from the direction of the Crab, the hadronic origin of TeV \gr s from it is considered to be less likely \citep{2021Sci...373..425L}. 
If the UHE emission from the Crab is hadronic, CRs should be accelerated in the pulsar’s magnetosphere or at the wind termination shock. 
An alternative scenario proposed by \citet{Cheng1990Crab} suggests that the CRs could be accelerated in the outer gap of the pulsar. 

We could only marginally constrain the hadronic fraction of the \gr\ flux for LHAASO J1849$-$0003 to $\lesssim 94\%$, with a power-law index of 1.99 \citep{Huang2021}. 

\section{Conclusions and outlook} \label{sec:conclude}
In this study, we have performed catalog and stacking searches on twelve LHAASO UHE \gr\ sources. We have found no significant neutrino emission, with the smallest post-trial p-value of 0.06 in PWN equal-weight stacking search. The resulting 90\% neutrino flux upper limits for an $E^{-2}$ spectrum are given. For LHAASO J1849-0003, LHAASO J0534+2202, and LHAASO J2226+6057, we are able to constrain the hadronic \gr\ emission tailored to the available spectral information of these sources. Joint fits between \gr s and neutrinos \citep{Abbasi2021jr} and physics-motivated lepto-hadronic modeling of the spectrum of each individual source may provide a better and more accurate limit. Future detectors should also improve our understanding of these potential PeVatrons. The planned IceCube-Gen2 detector \citep{ICgen2}, with an increased effective area relative to IceCube and improved angular resolution, is expected to become up to five times more sensitivity for Galactic neutrino sources compared to IceCube.

\begin{acknowledgments}
The IceCube collaboration acknowledges the significant contributions to this manuscript from Yu-Ling Chang, Kwok Lung Fan and Michael Larson. We also
acknowledge support from: USA {\textendash} U.S. National Science Foundation-Office of Polar Programs,
U.S. National Science Foundation-Physics Division,
U.S. National Science Foundation-EPSCoR,
Wisconsin Alumni Research Foundation,
Center for High Throughput Computing (CHTC) at the University of Wisconsin{\textendash}Madison,
Open Science Grid (OSG),
Advanced Cyberinfrastructure Coordination Ecosystem: Services {\&} Support (ACCESS),
Frontera computing project at the Texas Advanced Computing Center,
U.S. Department of Energy-National Energy Research Scientific Computing Center,
Particle astrophysics research computing center at the University of Maryland,
Institute for Cyber-Enabled Research at Michigan State University,
and Astroparticle physics computational facility at Marquette University;
Belgium {\textendash} Funds for Scientific Research (FRS-FNRS and FWO),
FWO Odysseus and Big Science programmes,
and Belgian Federal Science Policy Office (Belspo);
Germany {\textendash} Bundesministerium f{\"u}r Bildung und Forschung (BMBF),
Deutsche Forschungsgemeinschaft (DFG),
Helmholtz Alliance for Astroparticle Physics (HAP),
Initiative and Networking Fund of the Helmholtz Association,
Deutsches Elektronen Synchrotron (DESY),
and High Performance Computing cluster of the RWTH Aachen;
Sweden {\textendash} Swedish Research Council,
Swedish Polar Research Secretariat,
Swedish National Infrastructure for Computing (SNIC),
and Knut and Alice Wallenberg Foundation;
European Union {\textendash} EGI Advanced Computing for research;
Australia {\textendash} Australian Research Council;
Canada {\textendash} Natural Sciences and Engineering Research Council of Canada,
Calcul Qu{\'e}bec, Compute Ontario, Canada Foundation for Innovation, WestGrid, and Compute Canada;
Denmark {\textendash} Villum Fonden, Carlsberg Foundation, and European Commission;
New Zealand {\textendash} Marsden Fund;
Japan {\textendash} Japan Society for Promotion of Science (JSPS)
and Institute for Global Prominent Research (IGPR) of Chiba University;
Korea {\textendash} National Research Foundation of Korea (NRF);
Switzerland {\textendash} Swiss National Science Foundation (SNSF);
United Kingdom {\textendash} Department of Physics, University of Oxford.
\end{acknowledgments}
%



\appendix

\section{IceCube track-like neutrino dataset} \label{dataset}
\begin{figure}[h!]
\centering
\includegraphics[width=0.5\textwidth]{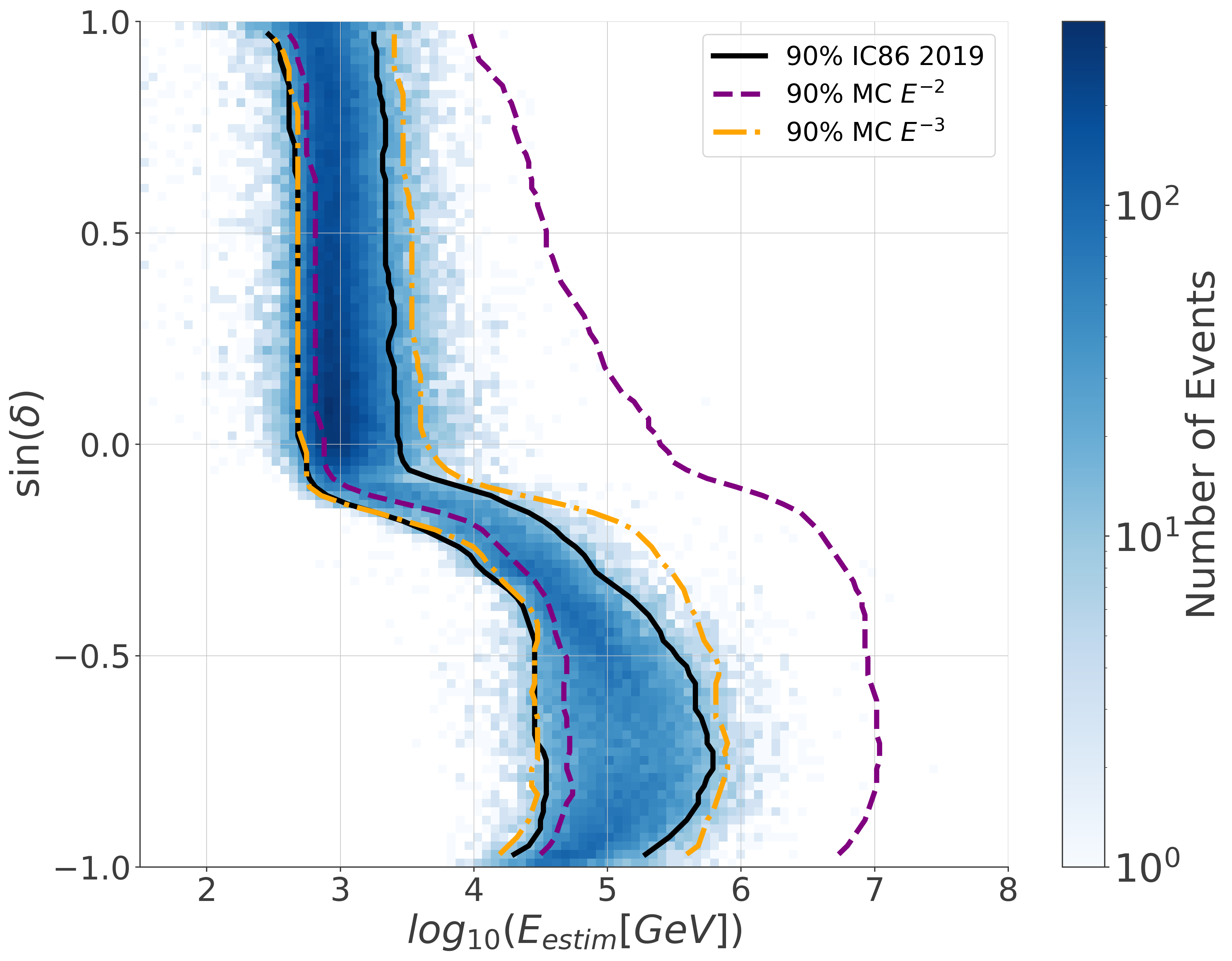}

  \caption{The 2D distribution of events in data taken in 2019 with full 86 strings as a function of
reconstructed declination and estimated energy. The 90\% energy range for the data is shown in the black solid line. Simulated
astrophysical signal Monte Carlo (MC) for an $E^{-2}$ and an $E^{-3}$
spectrum is shown in purple and orange
respectively as a guide for the relevant energy range of IceCube} \label{tab:fig4}
\end{figure}

\hyperref[tab:table4]{Table \ref*{tab:table4}} shows the event counts, livetime, start and end date of each detector configuration of the track-like neutrino dataset used in the analysis. \hyperref[tab:fig4]{Figure \ref*{tab:fig4}} shows the distribution of events as a function of reconstructed declination and estimated energy and also the 90\% energy range for data and MC signal with a spectral index of 2 and 3.

\begin{table}[h!]
\centering
\begin{tabular}{|ccccc|}
\hline
\multicolumn{1}{|c|}{Data} & \multicolumn{1}{c|}{Total number of events} & \multicolumn{1}{c|}{Total livetime (days)} & \multicolumn{1}{c|}{Start date} & Stop date \\ \hline
IC40                       & 36900                                      & 376                         & 6/4/2008                        & 20/5/2009                      \\
IC59                       & 107011                                     & 354                         & 20/5/2009                       & 31/5/2010                      \\
IC79                       & 101972                                     & 313                         & 1/6/2010                        & 13/5/2011                      \\
IC86                       & 1294410                                    & 3635                         & 13/5/2011                       & 22/8/2021       \\ \hline              
\end{tabular}
\caption{IceCube configuration, number of
events, livetime, start and end date.}
\label{tab:table4}
\end{table}

\section{Likelihood} \label{likelihood}
For a single-source analysis, the likelihood is 
\begin{equation}
L = \prod_i^N L_i(\vec{\theta},\vec{D_i}) = \prod_i^N(\frac{n_s}{N}S(\vec{\theta},\vec{D_i}) + \frac{N-n_s}{N}B(\vec{D_i})),
\end{equation}
where $L_i$ is the likelihood of the event i, $\vec{\theta}$ is the source properties including location, morphology, and spectrum. The parameter vector $\vec{D_i}$ stands for the reconstruction information of the event i, including reconstructed direction, energy, and estimated angular error. $N$ is the total number of events and $n_s$ is the best-fit number of signal neutrino. $S$ and $B$ are the signal probability distribution function (PDF) and background probability distribution function, each consisting of a product of a spatial likelihood and an energy likelihood. \\
Our test statistic (TS) is the log-likelihood ratio:
\begin{equation} 
TS = 2\sum_i^N log \left(\frac{L_i(\vec{\theta},\vec{D_i})}{L_i(n_s = 0,\vec{D_i}))}\right)\\ 
= 2\sum_i^N log\left( \frac{n_s}{N}(\frac{S(\vec{D_i},\vec{\theta})}{B(\vec{D_i})}-1)+1\right).
\end{equation}
The signal spatial likelihood of the point source is modeled as 2D Gaussian in the tangent plane which is valid for a sufficiently small angular extension
\begin{equation}    
S_{spatial}(\vec{r},\vec{r}_{\nu},\sigma)=\frac{1}{2\pi \sigma^2}e^{-\frac{(\vec{r}-\vec{r}_{\nu})^2}{2\sigma^2}}.
\end{equation}
Here, the parameters $\vec{r},\vec{r}_{\nu},\sigma$ correspond to the source location, reconstructed direction and estimated angular error. \\
The signal's spatial likelihood term for Gaussian extended sources is a modified version of the standard 2D Gaussian used for point sources.
\begin{equation}   
S_{spatial}(\vec{r}_{src},\vec{r},\sigma,\sigma_A)=\frac{1}{2\pi (\sigma^2+\sigma_A^2)}e^{{-\frac{(\vec{r}_{src}-\vec{r})^2}{2(\sigma^2+\sigma_A^2)}}},
\end{equation}
with the $\sigma_A$ being the extension of the source.
The background spatial likelihood is 
\begin{equation}    
B_{spatial}(\vec{r}_{events})=\frac{1}{2\pi }p(d),
\end{equation}
where $d$ is the reconstructed declination of the event and $p$ is the declination PDF of the background.
The energy-dependent part in the signal and background PDFs are modeled after the distributions of Monte Carlo and real data in terms of reconstructed energy and zenith.\\

For the stacking analysis, the likelihood is 
\begin{equation}
L =  \prod_i^N\frac{n_s}{N}(\sum_j^M w_j S(\vec{\theta_j},\vec{D_i}) + \frac{N-n_s}{N}B(\vec{D_i})),
\end{equation}
where $w_j$ is the product of the assumed source weight and energy-integrated effective area at the source declination. The weight $w_j$ is normalized as $\sum_j^M w_j = 1$ with M being the total number of sources\citep{abbasi2011time}. Hence the TS is 
\begin{equation} 
TS = 2\sum_i^N log\left( \frac{n_s}{N}(\frac{\sum_j^M w_j S(\vec{D_i},\vec{\theta_j})}{B(\vec{D_i})}-1)+1\right).
\end{equation}

\section{Connection between gamma-ray flux and neutrino flux} \label{convert}
In pp interactions, protons interact with nearby material and produce charged and neutral pions with a ratio of about 2 to 1 \citep{2014PhRvD..90b3010A}. Charged pion decay to neutrinos and other products, and the resulting neutrinos carry roughly 1/4 the energy of the pion. Neutral pion decay into two \gr s, each with 1/2 of the energy \citep{2014PhRvD..90b3010A}. On Earth, the flux observed is
\begin{equation}
E_\gamma J_\gamma(E_\gamma) \approx e^{\frac{-d}{\lambda_{\gamma\gamma}}} \frac{1}{3}\sum_{\nu_\alpha} E_\nu J_{\nu_\alpha}(E_\nu),
\end{equation}
where $J_\gamma(E_\gamma)$ and $J_{\nu_\alpha}(E_\nu)$ are the gamma-ray differential particle flux and neutrino differential particle flux for neutrino favor $\alpha$ . Assuming the absorption of gamma-rays is negligible and equal flux from each neutrino flavor due to oscillations, relationship between \gr and neutrino fluxes is then
\begin{equation}
J_{\nu_\mu}(E_{\nu_\mu}) = 2 J_\gamma(E_{\nu_\mu}/2).
\end{equation}



\bibliography{sample631}{}
\bibliographystyle{aasjournal}



\end{document}